\documentstyle[amsmath,graphics,graphicx,rotate,pifont,psfig]{aa}
\input{epsf}

\begin{document}
   \thesaurus{08         
              (02.08.1; 
               02.19.1; 
               02.20.1;
               09.03.1;
               09.11.1 )}
\title{The distribution of shock waves in driven supersonic turbulence}

\author{Michael D. Smith$^{1}$, Mordecai-Mark Mac Low$^{2}$ and Fabian Heitsch$^3$}
\offprints{M.D. Smith}

\institute{
$^1$ Armagh Observatory, College Hill, Armagh BT61 9DG, Northern Ireland\\
$^2$ Department of Astrophysics, American Museum of Natural History,
79th St. at Central Park West, New York, New York, 10024-5192, USA\\
$^3$ Max-Planck-Institut f\"ur Astronomie, K\"onigstuhl 17,
D-69117 Heidelberg, Germany\\
Internet: mds@star.arm.ac.uk, mordecai@amnh.org, heitsch@mpia-hd.mpg.de}
\date{Received; accepted}

\maketitle
\markboth{Smith et al: Driven Turbulence}{}

\begin{abstract} 

Supersonic turbulence generates distributions of shock waves.
Here, we analyse the shock waves in three-dimensional numerical 
simulations of uniformly driven supersonic turbulence,
with and without magnetohydrodynamics and self-gravity. 
We can identify the nature of the turbulence by measuring the
distribution of the shock strengths.

We find that uniformly  driven  turbulence possesses a power law 
distribution of fast shocks with the number of shocks inversely 
proportional to the square root of the shock jump speed. A tail of 
high speed shocks steeper than Gaussian results from the random
superposition of driving waves which decay rapidly. The energy is 
dissipated  by a small range of fast shocks. These results contrast 
with the exponential distribution and slow shock dissipation 
associated with decaying turbulence.
 
A strong magnetic field enhances the shock number transverse to the 
field direction at the expense of parallel shocks. A simulation with
self-gravity demonstrates the development of a  number of highly 
dissipative accretion shocks. Finally, we examine the dynamics to 
demonstrate how the power-law behaviour arises.

 
\end{abstract}

\section{Introduction}

Many structures we come across in the Universe have been shaped by fluid 
turbulence. In astronomy, we often observe high speed turbulence 
driven by supersonic ordered motions such as jets, supernova shocks,
galactic rotation and stellar winds (e.g. Franco \& Carrami\~nana 1999).
Although it has long been thought that astrophysical turbulence
provides the best opportunity to investigate supersonic turbulence, the 
lack of a theory has stunted our  attempts to understand the behaviour 
(von Hoerner 1962). Three dimensional high resolution numerical 
simulations now provide a method to make real progress. Driven 
turbulence is explored in this paper and the results compared to a 
sister study of decaying turbulence (Smith et al. 2000, hereafter 
Paper 1). Our aim here is to relate the type of turbulence to the 
properties of the shock waves.

We study here compressible turbulence without thermal conduction. 
No physical viscosity is modelled, but numerical viscosity remains 
present, and  an artificial viscosity determines the dissipation in 
regions of strong  convergence. Periodic boundary conditions were chosen 
for the finite difference ZEUS code simulations, fully described by 
Mac Low et al. (1998). Uniform three dimensional turbulence with an 
isothermal equation of state, an initially uniform magnetic field and 
periodic boundary conditions are investigated. The influence of 
self-gravity is also examined.

Our motivation here is to  provide the observer with a means of 
recognising the type of turbulence from the properties of the generated 
shock waves. The general energetics of decaying and driven hydrodynamic 
turbulence have already been computed by Mac Low et al (1998) and Mac 
Low (1999), respectively, using ZEUS, a second-order, Eulerian 
hydrocode. Mac Low et al (1999) concluded that, since turbulence which 
is left to decay dissipates rapidly under all conditions, the motions  
we observe in molecular clouds must be continuously driven. Klessen et 
al. (2000) extended the hydrodynamic results by calculating 
self-gravitating models. A smoothed particle hydrodynamics (SPH) code 
was also employed to confirm the results for both the decaying and 
driven cases (Mac Low et al 1998, Klessen et al. 2000).  Klessen et al. 
(2000) have provided the parameter scaling for applications to 
molecular clouds. Heitsch et al. (2000) present the magnetohydrodynamic 
extension to the self-gravitating case, and discuss the criteria for 
gravitational collapse.

Our immediate target is to derive the spectrum of shocks (the Shock 
Probability Distribution Function) generated by driven turbulence. With
this knowledge, we will proceed to predict the spectroscopic
properties in a following work. Observed individual bright, sharp 
features, such as arcs, filaments and  sheets, are often interpreted 
as shock layers within which particles are highly excited (e.g.
Eisl\"offel et al. 2000). Where unresolved,  the excitation  can still 
be explored quantitatively by employing spectroscopic methods. The 
gas excitation then depends on both the physics of shocks as well as the 
distribution of shock strengths.

Previous studies of compressible turbulence have concentrated on the 
density and velocity structure of the cold gas rather than the shocks
(e.g. Porter et al. 1994; Falgarone et al. 1994, V\'azquez-Semadeni 
et al. 1996, Padoan et al. 1998). This may be appropriate for the 
interpretation  of clouds since, although the Mach number is still 
high, the shock speeds are too low to produce bright features. The 
simulations analysed here are also being interpreted by Mac Low \& 
Ossenkopf (2000) in terms of density structure.

The method used to count shocks from grid-based simulations was 
developed in Paper 1. The one-dimensional counting procedure was 
verified through a comparison with full three-dimensional 
integrations of the dissipated energy in Paper 1. This method is
appropriate for a ZEUS-type finite difference code for which
shock transitions are spread out over a few zones. Here, we  also 
study the shock transitions in all three directions and display the
spatial distributions for the energy dissipated through the
artificial viscosity in the shocks. We first present the shock jump PDFs 
and provide analytical fits (Sect. 2). Magnetohydrodynamic 
simulations (Sect. 3) with Alf\'ven numbers of $A\,=\,5$ and $A\,=\,1$ are then   
explored. Note the definition of the  Alfv\'en Mach number 
$A = v_{\rm rms}/v_A$,  where the Alfv\'en speed $v_A = (B^2/4\pi\rho)^{1/2}$, 
and $v_{\rm rms}$ is the initial root mean square (rms) velocity.

Which shocks actually dissipate most of the energy? The shock power 
rather than number density will determine the spectral region in which 
the system can be best observed. Hence, we determine the energy dissipated 
as a function of the shock speed in Sect. 4. A large-scale simulation which 
included self-gravity is then analysed in Sect. 5. We next interpret the results 
in terms of the dynamical models (Sect. 6).

\section{Hydrodynamic  Turbulence}
 
\subsection{Model description}

We first explore driven hydrodynamic turbulence. The three-dimensional 
numerical simulations on grids with $128^3$ and $256^3$ zones and 
periodic boundary conditions were initialized with an rms Mach number of
$M=5$. The initial density is uniform and the initial velocity 
perturbations were drawn from a Gaussian random field, as described by 
Mac Low (1999). The power spectrum of the perturbations is flat and 
limited to small wavenumber ranges,  $k_{\rm min} < k <  k_{\rm max}$, 
with  $k_{\rm min} = 7$ and  $k_{\rm max} = 8$ in Fig.\,\ref{hydro}. 
The uniform driver is a simple constant rate of energy input with the  
distribution pattern fixed.

The simulations employ a box of length $2L$ and a unit of speed, $u$. The 
gas is isothermal with a sound speed of $c_s = 0.1\,u$. Hence the  sound 
crossing time is 20\,L/u. We take the time unit as $\tau = L/u$. Assuming a 
mass m to be contained initially in a cube of side $L$, the dissipated power
is given in units of $\dot E_o = mu^2/\tau$. 

\subsection{Shock number distribution}         

We  calculate the one dimensional shock jump function, as discussed
and justified in Paper 1. This is the number distribution  of the total 
jump in speed across each converging region along a specific direction. 
This is written as $dN/dv_j$ where $v_j$ is the sum of the (negative) 
velocity gradients (i.e. $\Sigma\left[-{\delta}v_x\right]$ across a 
region being compressed in the x-direction). We employ the jump Mach 
number in the x-direction $M_j = v_j/c_s$ rather than $v_j$ since this 
is the parameter relevant to the dynamics. Thus, each bounded region of 
convergence in the x-direction counts as a single shock and the total 
jump in $M_j$ across this region is related to its strength. 

Numerically, we scan over the whole  simulation grid 
(x,y,z), recording  each shock jump through
\begin{equation}
    M_j =  \sum_{x=x_i}^{x=x_f} ({\Delta}v_x/c_s)
\end{equation}
with the condition that ${\Delta}v_x < 0$ in the range $x_i \le x < x_f$.
This is then binned  as a single shock element. The shock number 
distribution $dN/dM_j$ is obviously dimensionless. 

Note that the shock number depends on the grid size. We find that on 
doubling the grid size, the shock number increases by a factor of four.
This implies that the total shock front {\em area} in units of $L^2$ remains
roughly constant. This holds for both driven and decaying turbulence and is 
an indication that the numerical resolution is sufficient to capture
the vast majority of shocks. 

\subsection{Steady state description}                 

\begin{figure}
  \begin{center}
    \leavevmode
    \psfig{file=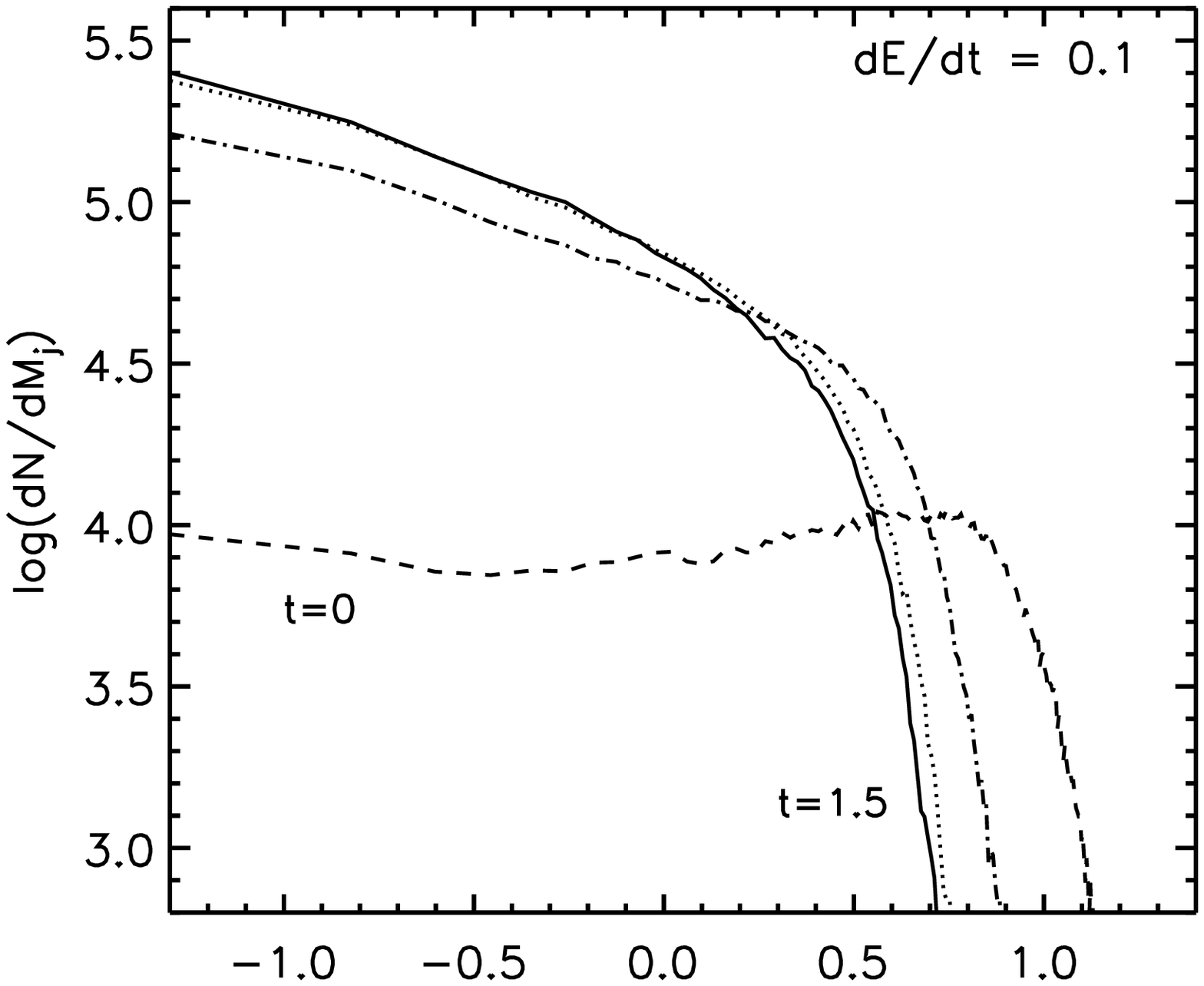,width=0.48\textwidth,angle=0}
    \psfig{file=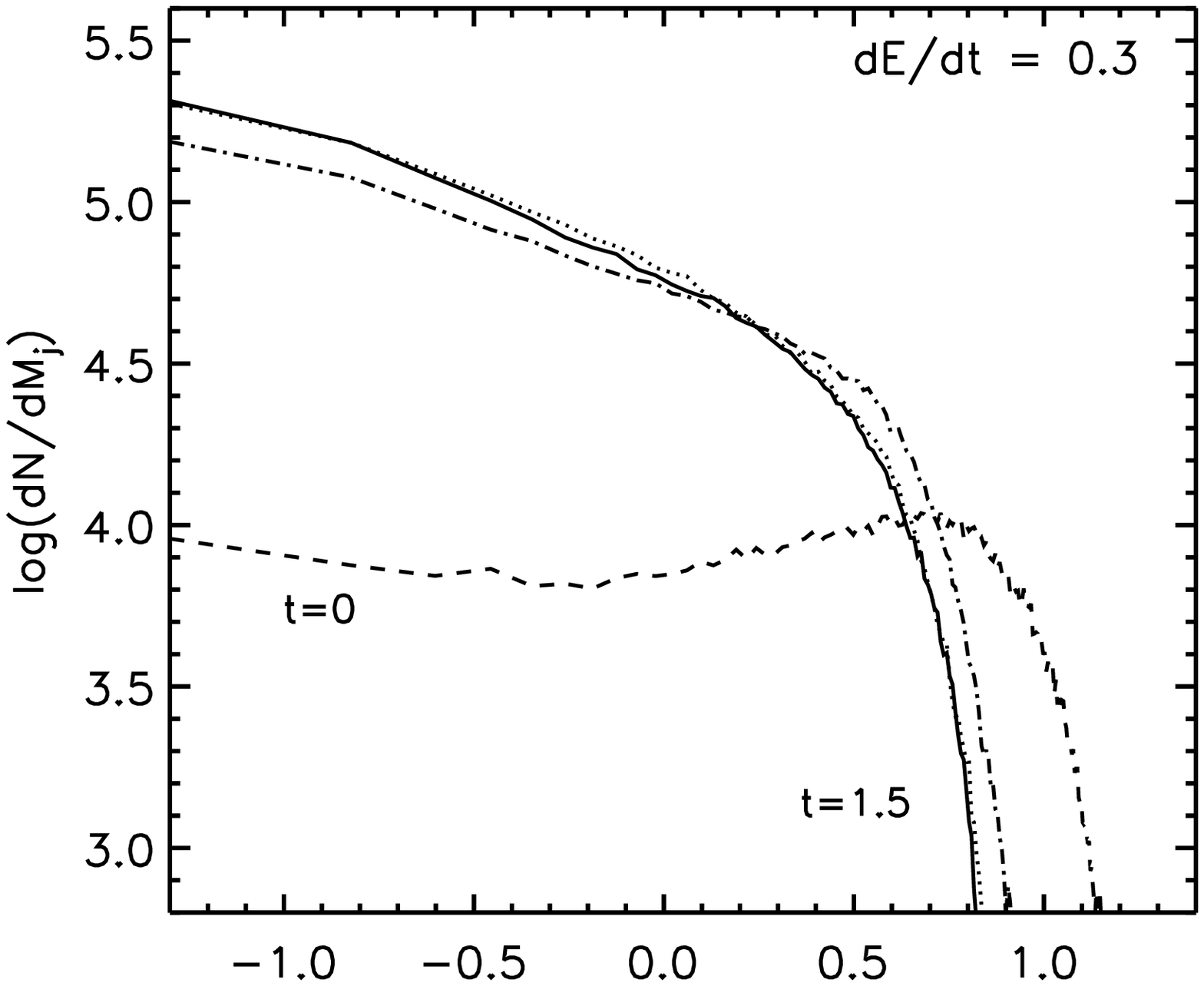,width=0.48\textwidth,angle=0}
    \psfig{file=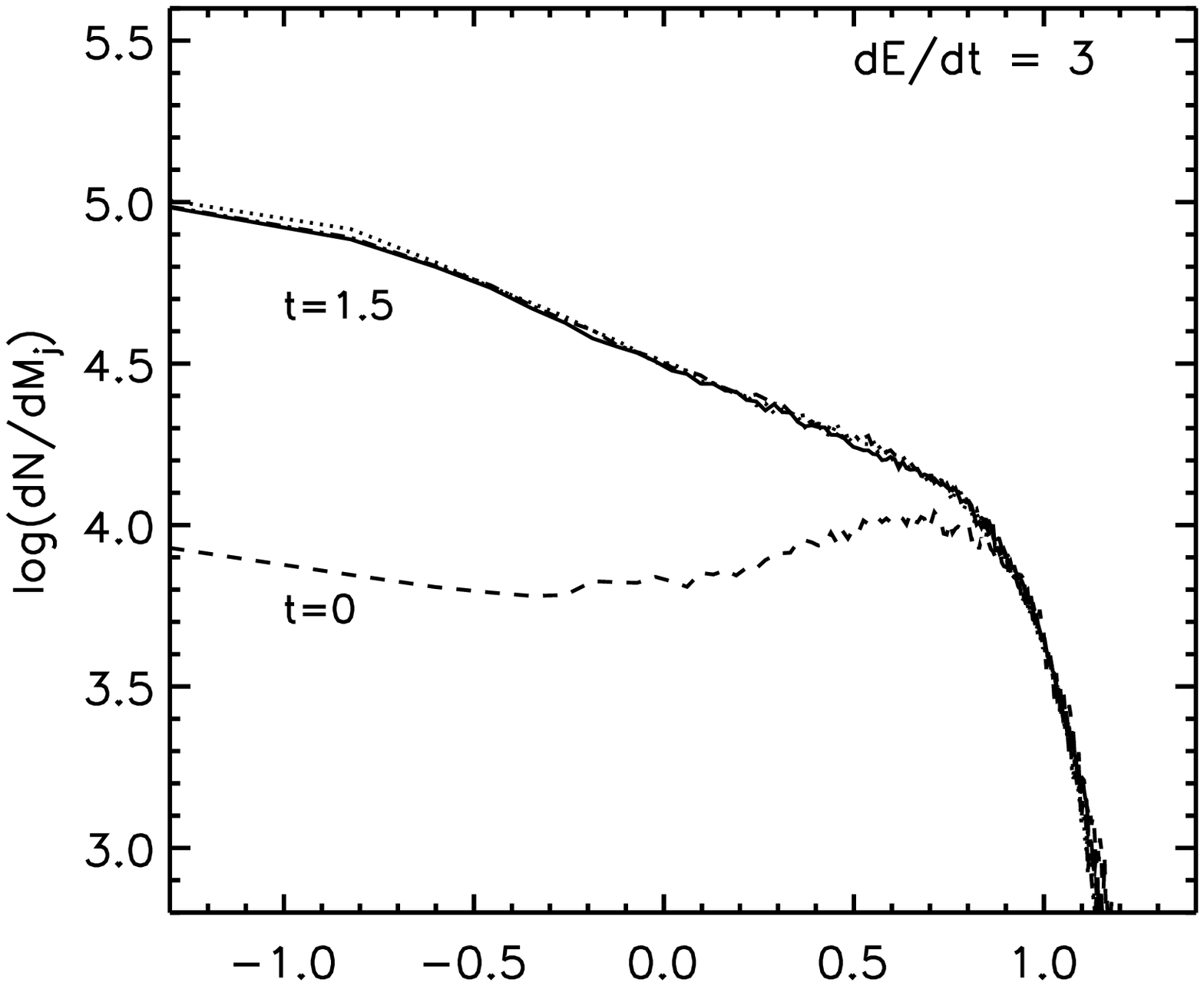,width=0.48\textwidth,angle=0}
    \psfig{file=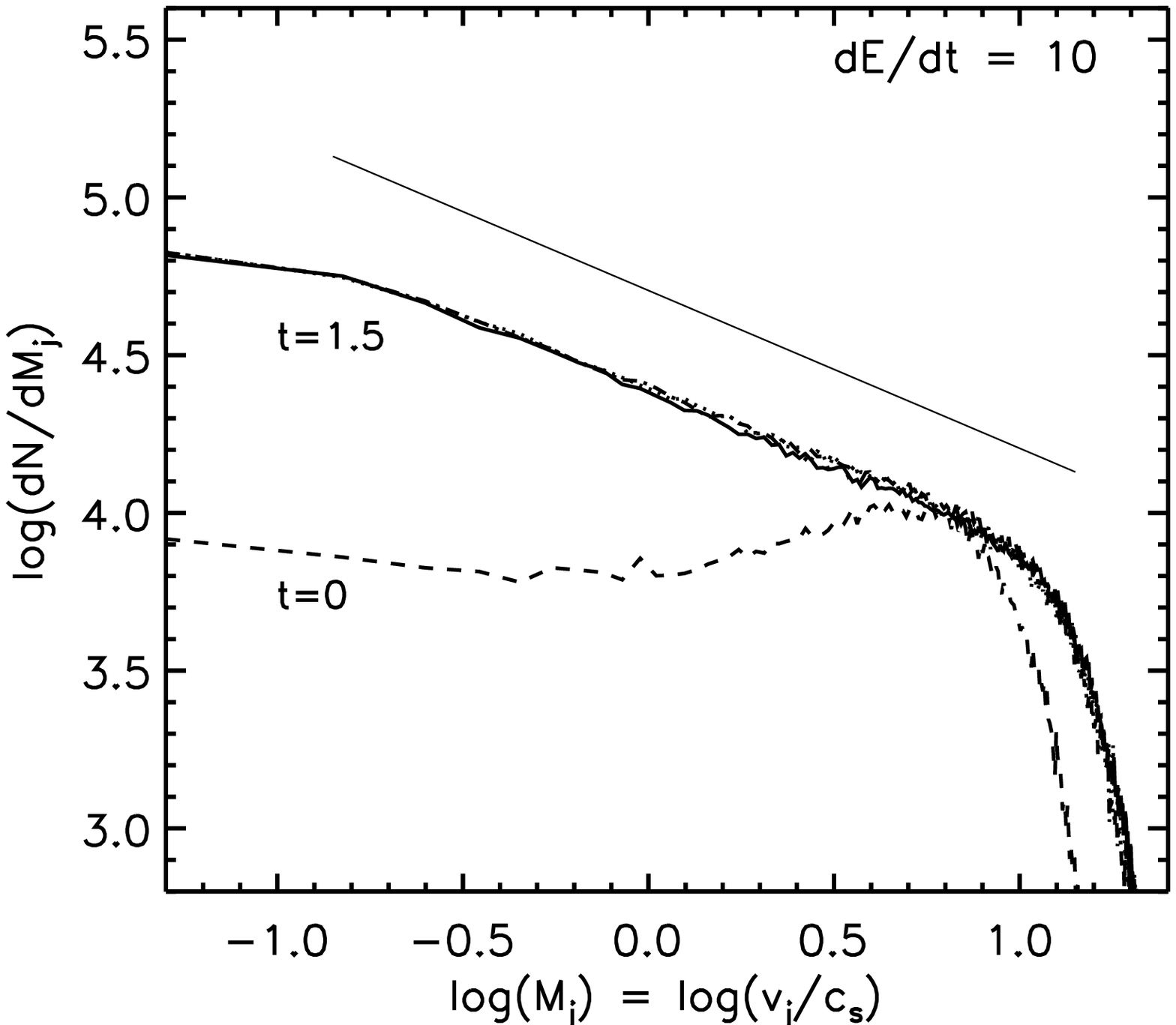,width=0.48\textwidth,angle=0}
\caption{Four simulations of driven hydrodynamic turbulence 
with an increasing rate of energy 
input from top to bottom. The distribution of  jumps are shown at four 
times for each simulation, from t=0 (dashed), t=0.5 (dot-dash), t=1.0 
(dotted) to t=1.5 (solid). The thin straight line in the lower panel
represents an inverse square root law.}  
\label{hydro}
  \end{center}
\end{figure}

The random Gaussian field at t\,=\,0  rapidly transforms into a shock field
(Fig\,\ref{hydro}). As can be seen, the shock distribution approaches a steady state, 
reached by t\,=\,1. The driving energy determines the break in
the distribution and the maximum shock
speed, but does not influence the {\em distribution} of shock speeds.
  
\begin{figure}[hbt]
  \begin{center}
    \leavevmode
    \psfig{file=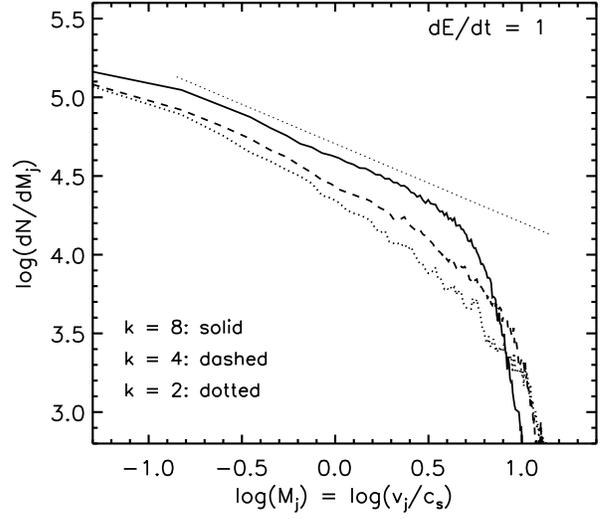,width=0.60\textwidth,angle=0}
\caption{The shock distribution depends on
the wavenumber of the energy input of the driven hydrodynamic turbulence. 
The straight dotted line has the slope given by Eq.\,(\ref{eqnnumb}).}   
\label{wavenumber}
  \end{center}
\end{figure}

The driving wavenumber influences the steady state as shown
in Fig.\,\ref{wavenumber}.
There is a moderate dependence of the shock number on the
wavenumber, especially at high Mach numbers. This may be partly
due to the time involved for the regenerated low wavenumber modes to
steepen. Some of the longer wave modes may be damped by interactions with
the turbulent field.

A power law distribution of shock velocities is uncovered.
Note that for high Mach numbers, statistically,
we can equate the shock Mach number to the jump Mach number to a
good approximation. 
We find, as shown by the indicated line in the lower
box of  Fig.\,\ref{hydro}, an inverse square-root law 
${dN}/{dM_j} \propto M_j^{-0.5}$. In detail, we find a fit of the
form
\begin{equation}
\frac{dN}{dM_j} \sim 1.4\,10^{4}\,k^{0.5}\,\dot E^{-0.2}\,M_j^{-0.5}
\label{eqnnumb}
\end{equation}
over the power law sections. The break in the power law is
found to occur at $M_j^{max}$ given by
\begin{equation}
M_j^{max} \sim 7.6\,k^{0.2}\,\dot E^{0.4}.
\label{eqnmax}
\end{equation}
The k-dependence for $M_j^{max}$ has been estimated by inspecting 
the energy dissipation diagrams below. The indicated value is
consistent with that discernable in Fig.\,\ref{wavenumber}.
 
Integrating Eq.\,(\ref{eqnnumb}), using the limit Eq.\,(\ref{eqnmax}), yields
the remarkable result that the number of shocks is a constant: 
$N \sim 178,000$ for $k\,=\,8$. That is, when the energy input is 
low, there
are many more weak shocks. As the energy is increased, the 
number of shocks does not increase. Rather, the shock
strengths increase. Hence, the number of shocks reaches
a saturation level of about 180,000 on the 128$^3$ grids.

Hence, the total shock surface area is  $N \propto k^{0.4}$. This is 
confirmed directly from the shock counts. We find that the number of 
jumps does depend weakly on the chosen value for the minimum
convergence. We take the case $dE/dt = 10$ for
illustrative numbers. Taking all converging flow regions, yields
181,000 shocks with 5.17 zones per shock. This gives 46\% of
the volume occupied by converging regions. Taking instead a minimum
convergence of $0.5\,c_s$ as a criteria which 
relates to a steepened wave balanced by artificial viscosity,
yields 161,000 regions with an average of 3.95 zones. That is,
almost 90\% of the converging regions are indeed associated with
steepened waves, which occupy a constant 30\% of the volume.
This number density was also found in decaying turbulence:
even though the shocks may weaken and interact, the
total number is conserved, and the total shock area
is independent of the grid size (Paper 1).

\begin{figure}[hbt]
  \begin{center}
    \leavevmode
    \psfig{file=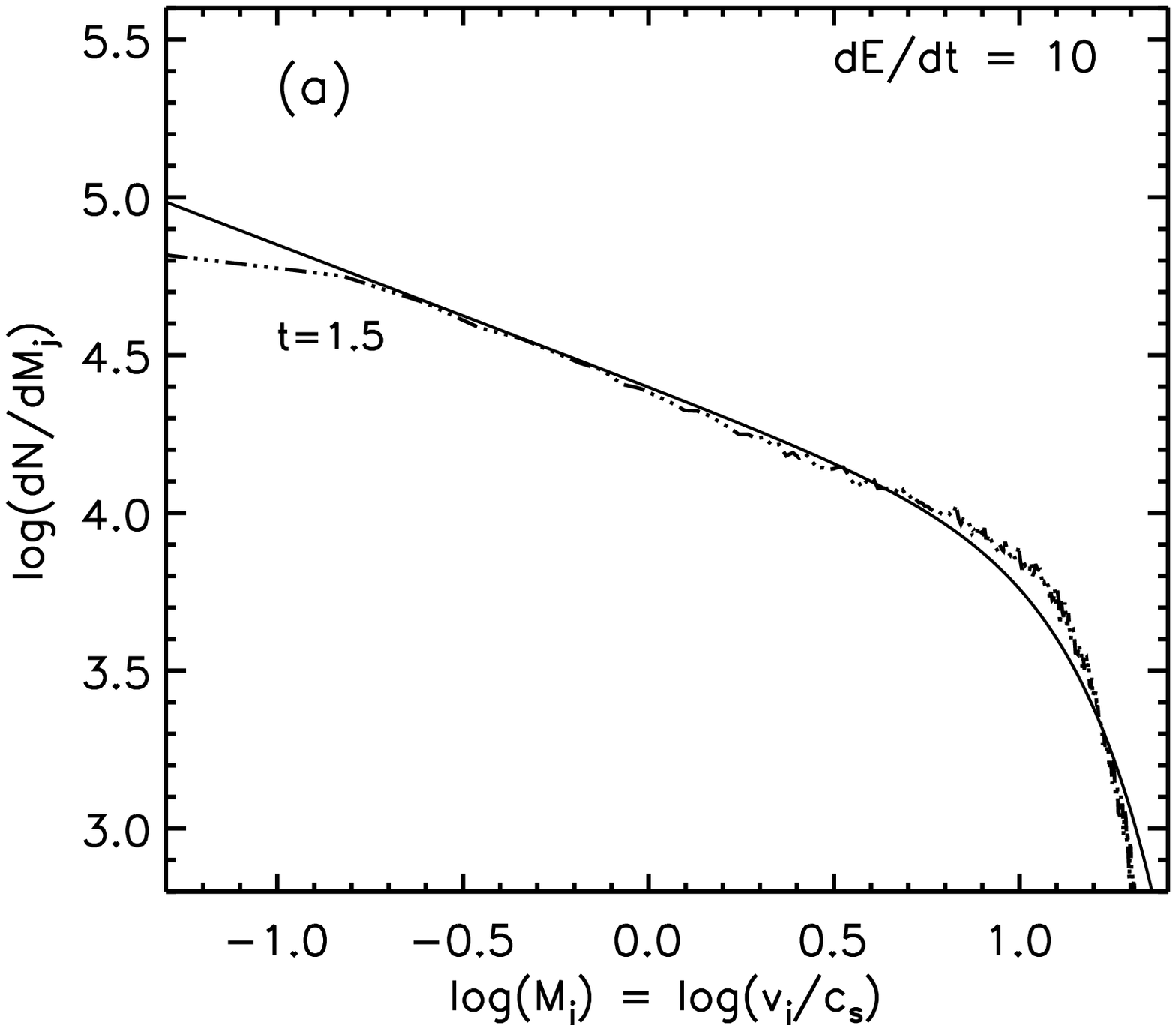,width=0.60\textwidth,angle=0}
    \psfig{file=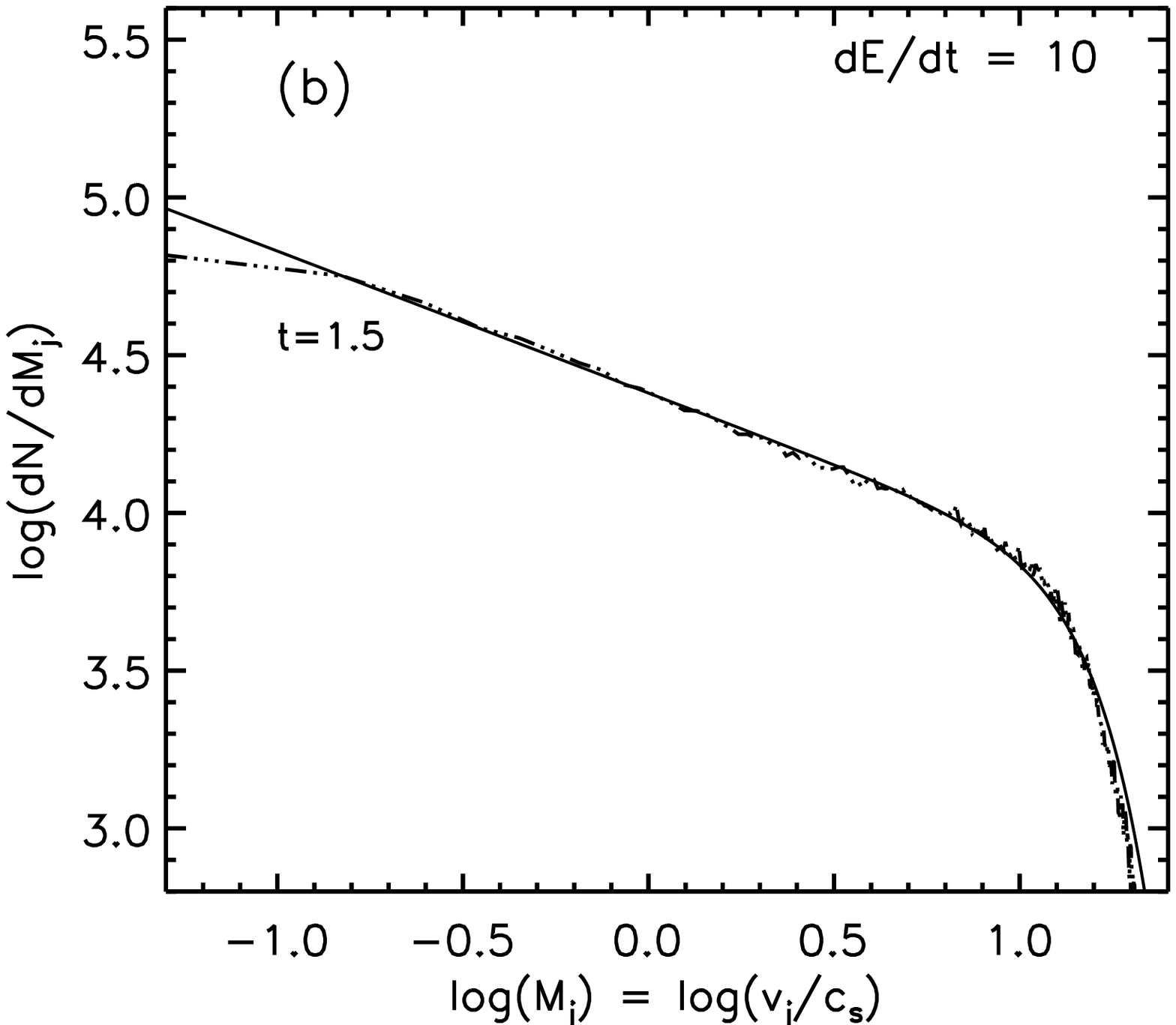,width=0.60\textwidth,angle=0}
\caption{Power law fits  combined with a (a) Gaussian tail and (b)
a steep exponential tail of the form $\exp \left[-(M/16.6)^3\right]$ 
to the shock number distribution for the case $k = 8$ and $\dot E = 10$.}   
\label{tail}
  \end{center}
\end{figure}

Many predictions and analytical fits to simulations have been
published for the velocity gradients within turbulent flows.
Here, for supersonic turbulence,
we find a  Gaussian high-speed tail to the shock distribution
provides  a rough fit. Figure\,\ref{tail}
displays a combined power-law and Gaussian fit of the form
\begin{equation}
\frac{dN}{dM_j} \sim 1.4\,10^{4}\,k^{0.5}\,\dot E^{-0.2}\,M_j^{-0.45}\,
                 \exp \left[-(M_j/15.1)^2\right]
\label{eqntail}
\end{equation}
for one case. To test if this
is due to the initial Gaussian field we have also
run a simulation with an exponential driver and found that a similar tail
is still present. Hence the near-Gaussian could be due to the superposition 
of the initially randomly-located forcing waves, as expected from the Central
Limit Theorem. The best fit we find , however, involves a tail of the form
$\exp (-M^{3})$,
\begin{equation}
\frac{dN}{dM_j} \sim 1.4\,10^{4}\,k^{0.5}\,\dot E^{-0.2}\,M_j^{-0.45}\,
                 \exp \left[-(M_j/16.6)^3\right].
\label{eqntail}
\end{equation}
Such steep tails have indeed been commonly found despite Gaussian
driving, thought to be due to the more rapid decay
of the faster shocks (e.g. Gotoh \& Kraichnan 1998).

\section{Magnetohydrodynamic  Turbulence}

\begin{figure}[hbt]
  \begin{center}
    \leavevmode
    \psfig{file=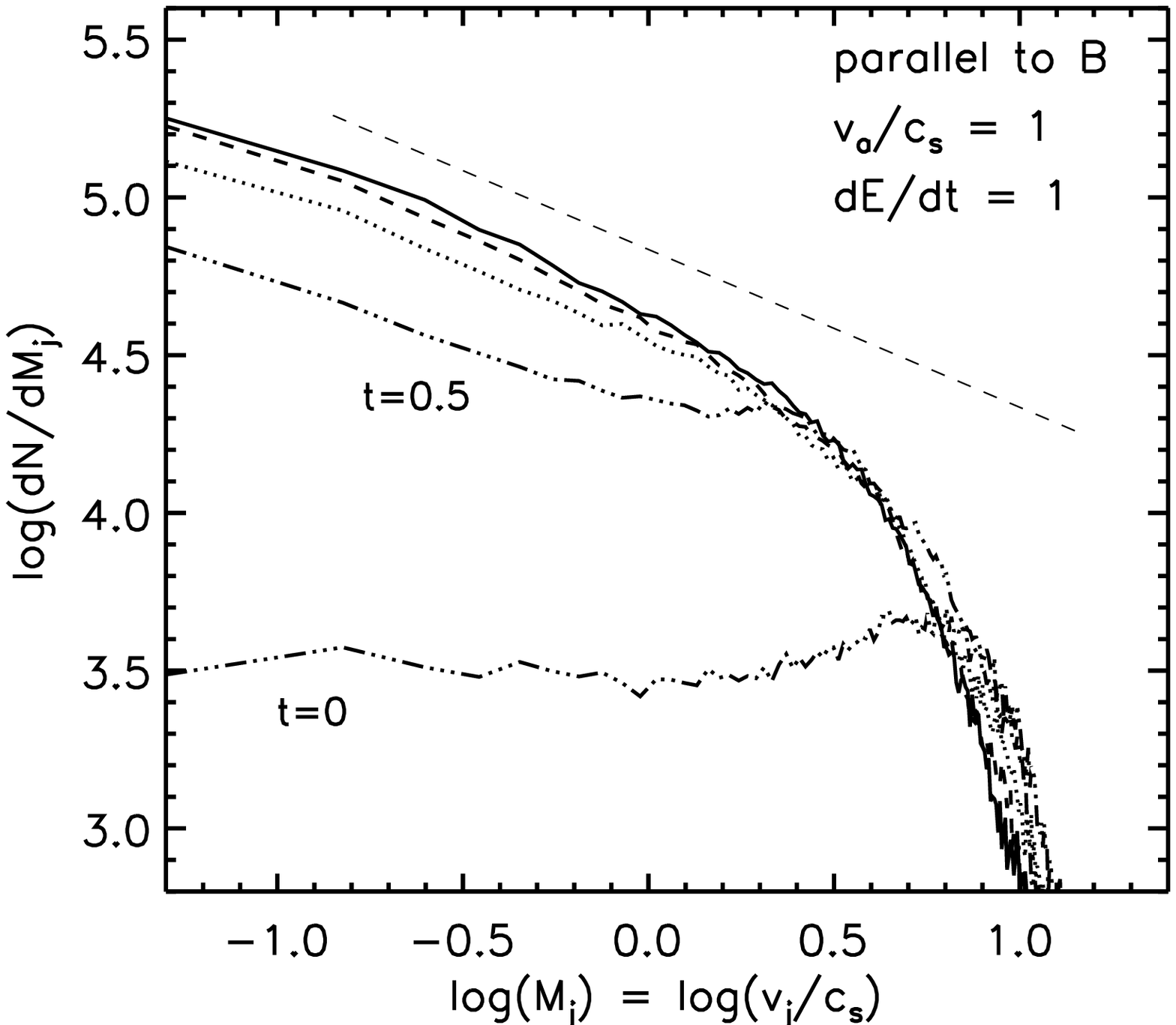,width=0.50\textwidth,angle=0}
    \psfig{file=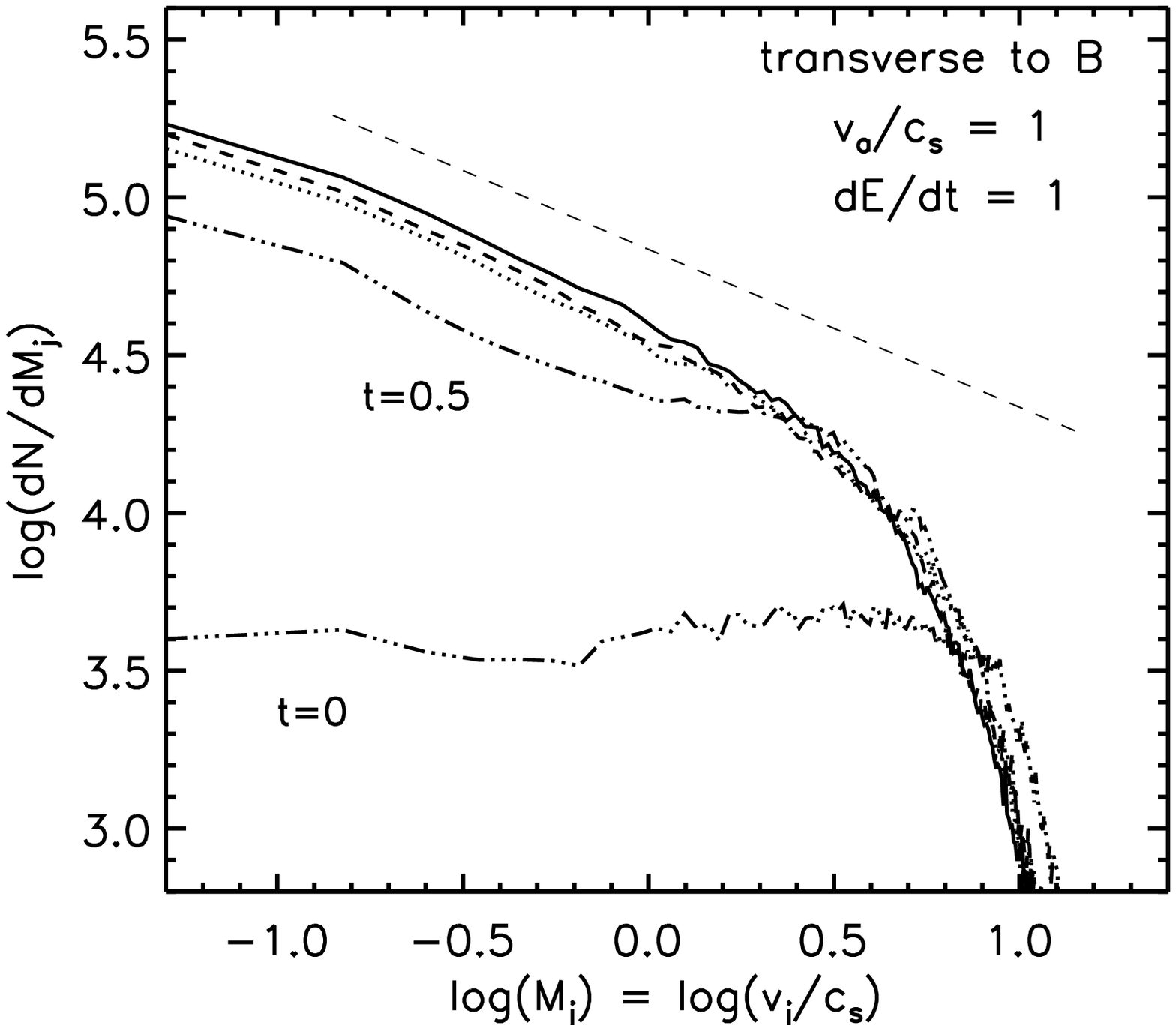,width=0.50\textwidth,angle=0}
\caption{The distribution of shocks in driven MHD turbulence with a 
'weak' magnetic field. The shock numbers are shown here parallel and 
transverse to the field, at times t=0, t=0.5 (dot-dash), t=1.0 (dotted) 
to t=1.5 (dashed) and the time t=3.0 (solid), by which time a statistical 
steady state has been reached. The driving wavenumber is $k = 4$. The 
straight dashed line represents an inverse square root power law.}   
\label{maglow}
  \end{center}
\end{figure}

We introduce a uniform  magnetic field into the initial configuration.
A weak field, in which the Alfv\'en and sound speeds are equal, has
little overall influence (Fig.\,\ref{maglow}). The shock distribution 
is roughly isotropic and the shock number is not significantly altered from
the equivalent hydrodynamic simulation. The power law section is not so well
defined for the shock distribution parallel to the field.

\begin{figure}[hbt]
  \begin{center}
    \leavevmode
    \psfig{file=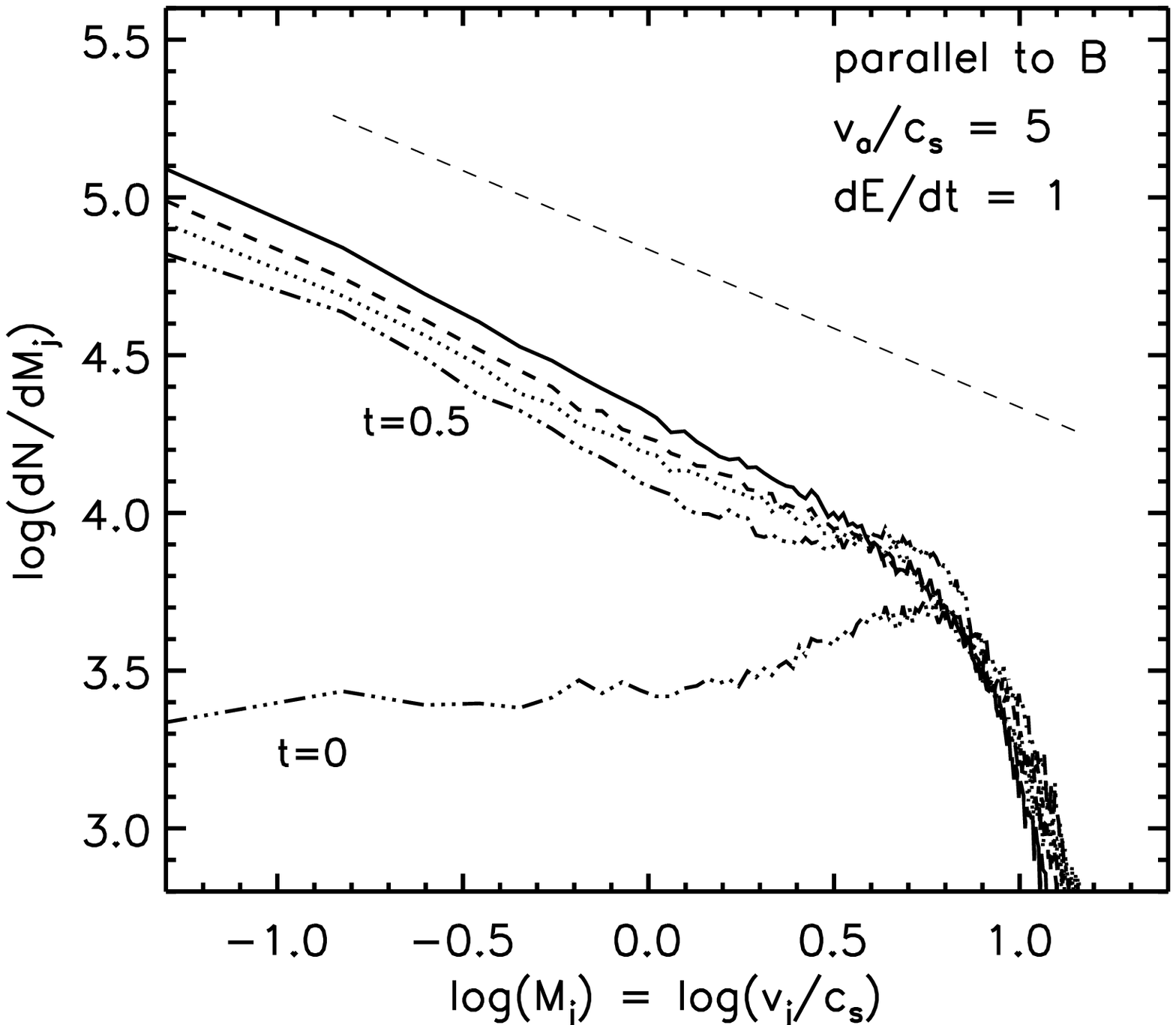,width=0.50\textwidth,angle=0}
    \psfig{file=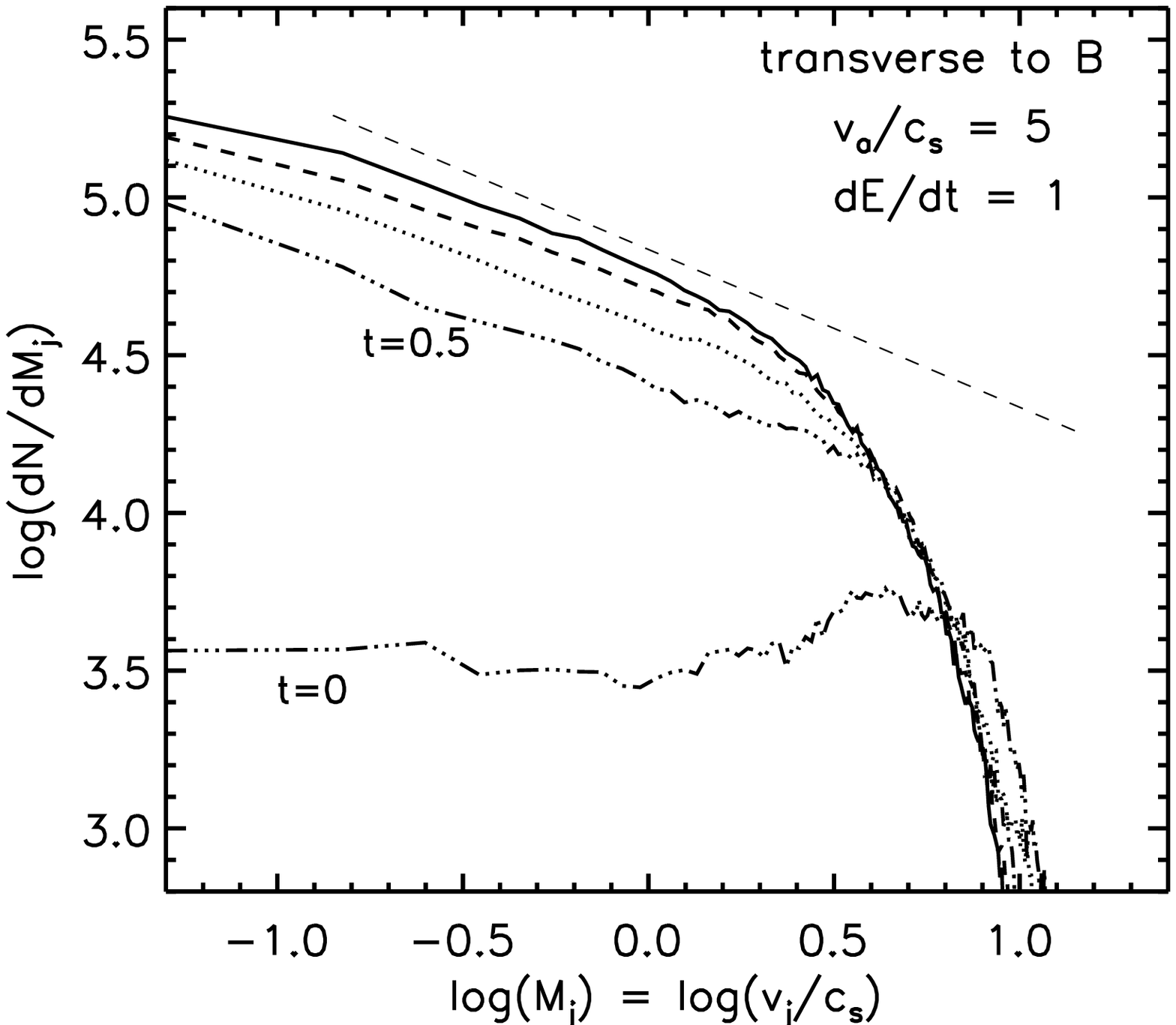,width=0.50\textwidth,angle=0}
\caption{The distribution of shocks in driven MHD turbulence with a 
strong magnetic field. The shock numbers are shown here parallel and 
transverse to the field, at times t=0, t=0.5 (dot-dash), t=1.0 (dotted) 
to t=1.5 (dashed) and the time t=3.0 (solid), by which time a 
statistical steady state has been reached.}   
\label{maghigh}
  \end{center}
\end{figure}

A strong field introduces a strong anisotropy (Fig.\,\ref{maghigh}). 
With a field such that the Alfv\'en speed equals the initial rms
speed, {\em the waves transverse to the field dominate}. There are 
$\sim 2-3$ times more waves in the transverse direction for a
given jump speed in each direction. The inverse square root power-law
rule is again closely obeyed. We call these waves, rather than shocks,
since the high Alfv\'en speed   implies that a high fraction may be
fast magnetosonic waves. The average number of zones, however,
measured for each jump is only 4.7 (parallel) and 7.4 (transverse).
This compares to an average of 5.4 for the hydrodynamic flow (and 23.7
zones for the initial Gaussian with k=4). Note these refer to the
complete shock, not just the 2-3 zones across which the jump is
highly non-linear and across which numerical viscosity is strong);
there usually exists one or two zones on each side of the main 
jump across which the velocity joins smoothly onto the surrounding flow 
without oscillations). Hence, the zone measurements favour the 
interpretation that the dissipation is being carried out in 
short-wavelength non-linear magnetosonic waves.

A very similar difference between parallel and transverse shock 
numbers is found in the case of decaying turbulence (Paper 1). 
It is clear that the magnetic pressure is not damping the shock 
waves. In contrast, the extra magnetic energy which becomes tied 
up in the waves also helps maintain them.


\section{Energy dissipation}

The increasing number of strong shocks with increasing  wavenumber is
qualitatively consistent with the finding of Mac Low (1999) that the total 
energy dissipation rate increases with the driving wavenumber. To discuss 
the energy dissipation rate, however, we must first also consider the 
density $\rho$ within each shock. We have then calculated the power 
dissipated by artificial viscosity within each shock front, as described 
in Paper 1. This yields the power dissipated per unit shock speed. We 
actually calculate in this section the component of the dissipation along 
the x-axis and the jump Mach number along the x-axis which, given the 
statistics,  represents the true three dimensional `power distribution 
function'.  Numerically, over the whole simulation grid (x,y,z), we identify  
each compact range  $x_i \le x < x_f$ along the x-axis for which 
${\Delta}v_x = v_{x+1}-v_x < 0$. For each jump we then find 
 the energy dissipated by artificial viscosity as
\begin{equation}
       P_j =  \frac{C}{L^2} \sum_{x=x_i}^{x=x_f} \rho_x({\Delta}v_x)^3,
\end{equation}
where C  measures the number of zones over which artificial viscosity 
will spread a shock. This is then binned  as a single shock element. The 
shock number distribution $dP/dM_j$ is multiplied by three to account for 
the energy dissipated in each of the three dimensions. Further details of 
the method can be found in Paper 1, where its reliability was also verified. 
 
\begin{figure*}[hbt]
  \begin{center}
    \leavevmode
    \psfig{file=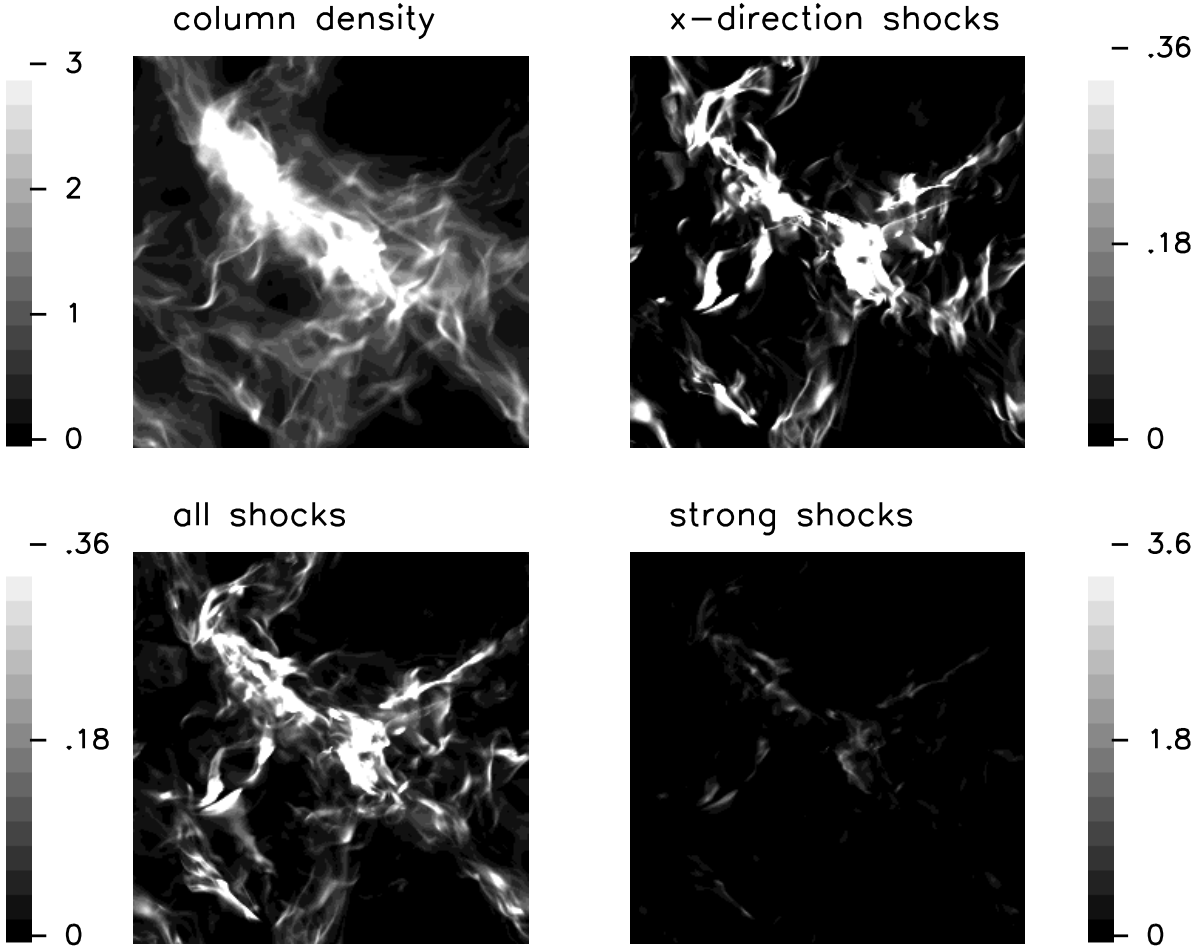,width=0.94\textwidth,angle=0}
\caption{Maps of the column and power dissipated from driven
turbulence integrated through the cube along the z-direction.
The run is D1 of Klessen et al.
2000, at time t\,=\,2, just before self-gravity was switched on. The 
long wavelength driving (k$_{\rm max} = 2$) generates a large cloud structure
visible in the column density, the density projected onto the x-y plane.
The energy dissipated in the shocks (1) just in the x-direction,
(2) in all directions and (3) in just the strongest shocks, are
displayed as indicated. The column density is expressed relative to the
average (initial) value, and the power loss per image pixel has been amplified
by a factor of 128$^2$.}   
\label{image-dr}
  \end{center}
\end{figure*}

The surface brightness of the shocks, as well as the column density of all 
the gas within the cube, is displayed in Fig.\,\ref{image-dr} for a pure
hydrodynamic driven example. Note that these are not slices, but we
have integrated through the z-direction. Much of the gas has been swept up
into a flattened cloud.
Many shocks appear sharper when
only the x-direction is accounted for since this emphasizes
the shocks transverse to the line of sight.  

Dissipating shocks and  the density distribution do not correlate
very well, although the main cloud contains the main
elongated region of dissipation. The  most
powerful shocks are mainly within the cloud.

\begin{figure}[hbt]
  \begin{center}
    \psfig{file=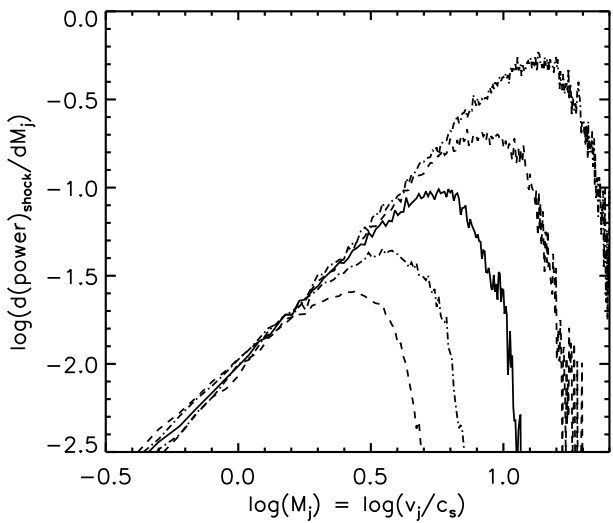,width=0.60\textwidth,angle=0}
\caption{The rate of energy dissipation from shocks with jump
Mach number M$_j$. The log-log plot demonstrates that power law
relation. The 5 curves correspond to the simulations shown in
Fig.\,\ref{hydro} with increasing input powers 
dE/dt = (0.1, 0.3, 1, 3, 10).}  
\label{energy}
  \end{center}
\end{figure}

\begin{figure}[hbt]
  \begin{center}
    \psfig{file=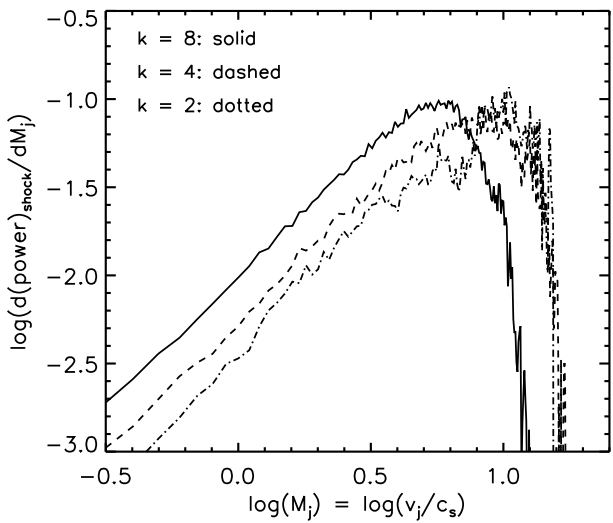,width=0.60\textwidth,angle=0}
\caption{The distributions of the rate of energy dissipation for the
three indicated maximum wavenumbers, corresponding to the number
distributions shown in Fig.\,\ref{wavenumber}.}   
\label{energyk}
  \end{center}
\end{figure}

We find that the energy dissipation rate also takes on a power law
$M_j$ dependence (Fig.\,\ref{energy}). In contrast to the number 
distribution, it is independent of the driving power input. There
is also a wavenumber dependence, as shown in Fig.\,\ref{energyk}.

The total energy dissipated in the power law section (summing
the losses for each direction) can be written
\begin{equation}
   \frac{d P}{dM_j} \sim 1.05\,10^{-2}\,k^{0.5}\,M_j^{1.5}.
\label{eqnpower}
\end{equation}
The error in the power law index of 1.5 is $\sim 0.12$. The break in 
the power law is found to occur at $M_j^{max}$ as given by 
Eq.\,(\ref{eqnmax}).
Integrating Eq.\,(\ref{eqnpower}), we obtain Eq.\,(\ref{eqnmax}). Thus we have 
acquired a self-consistent mathematical description.

\section{Gravitational Collapse}

\begin{figure}[hbt]
  \begin{center}
    \leavevmode
    \psfig{file=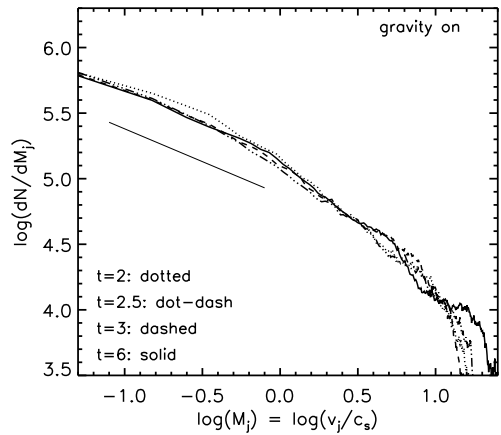,width=0.50\textwidth,angle=0}
\caption{The distribution of shocks 
in driven turbulence with gravity. The initial rms speed
is 9.9$c_s$, $\dot E = 0.4$,  wavenumber $k = 2$ and
the density is chosen to be 0.125 (total mass of unity in the
periodic box). The gravitational constant is set to unity. The
gravity is switched on after the driven turbulence is
well established at time $t\,=\,2$. The turbulent Jeans mass
is 3.2 (see Klessen et al. 2000).}   
\label{selfg}
  \end{center}
\end{figure}

Can we distinguish a collapsing self-gravitating cloud from a 
turbulent region within which the Jeans mass is not reached?
We here take a simulation of driven hydrodynamic turbulence in which
self-gravity is switched on after a steady driven state has been reached
(Model D2 from Klessen et al. 2000).
The particular parameters chosen are shown in the caption to
Fig.\,\ref{selfg}. 

The number and distribution of shock speeds is not particularly
different from the equivalent non-gravity numerical experiment.
The power-law section is limited, consistent with other low
wavenumber simulations (see Fig.\,\ref{wavenumber}).

\begin{figure}
  \begin{center}
    \leavevmode
    \psfig{file=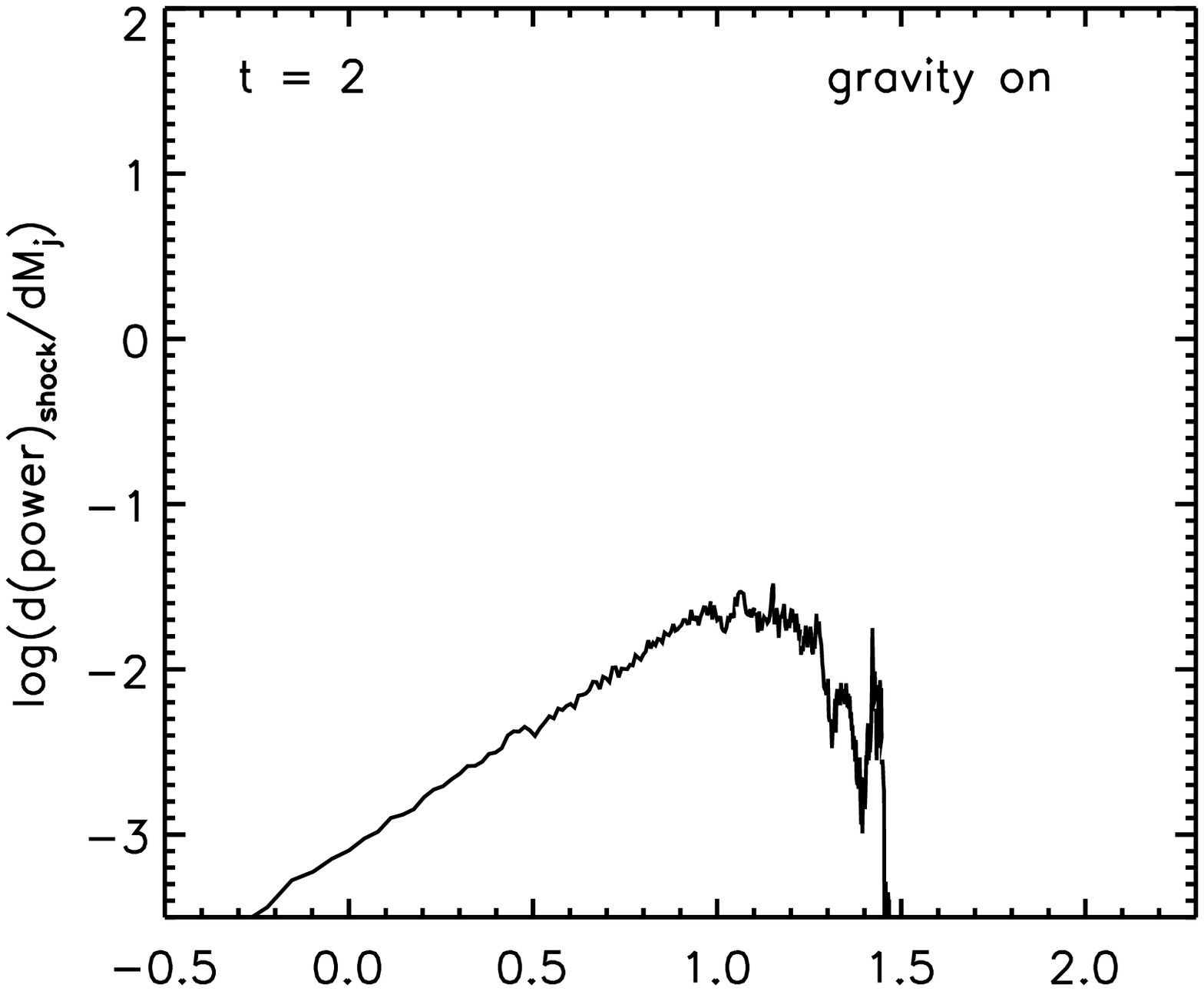,width=0.50\textwidth,angle=0}
    \psfig{file=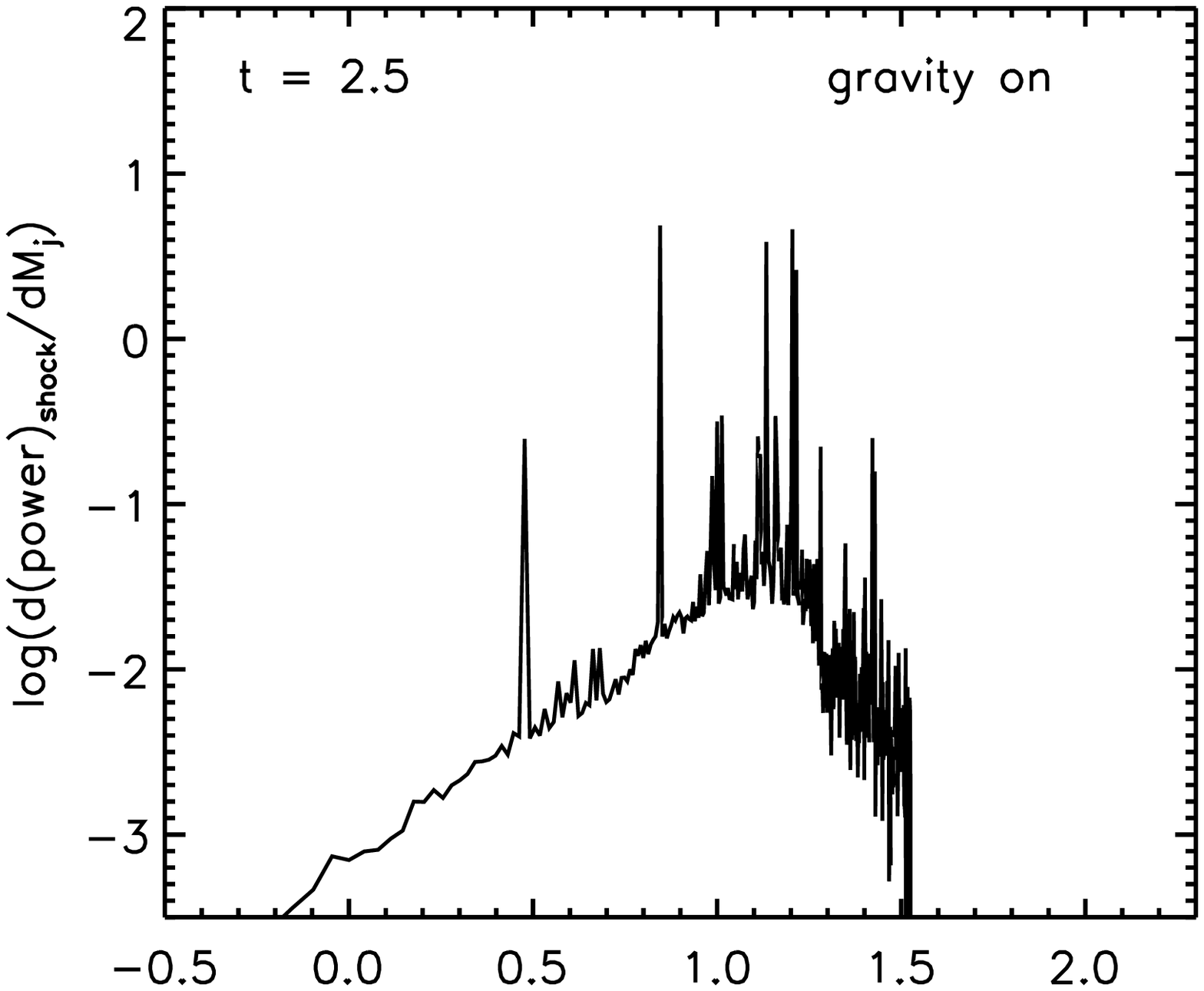,width=0.50\textwidth,angle=0}
    \psfig{file=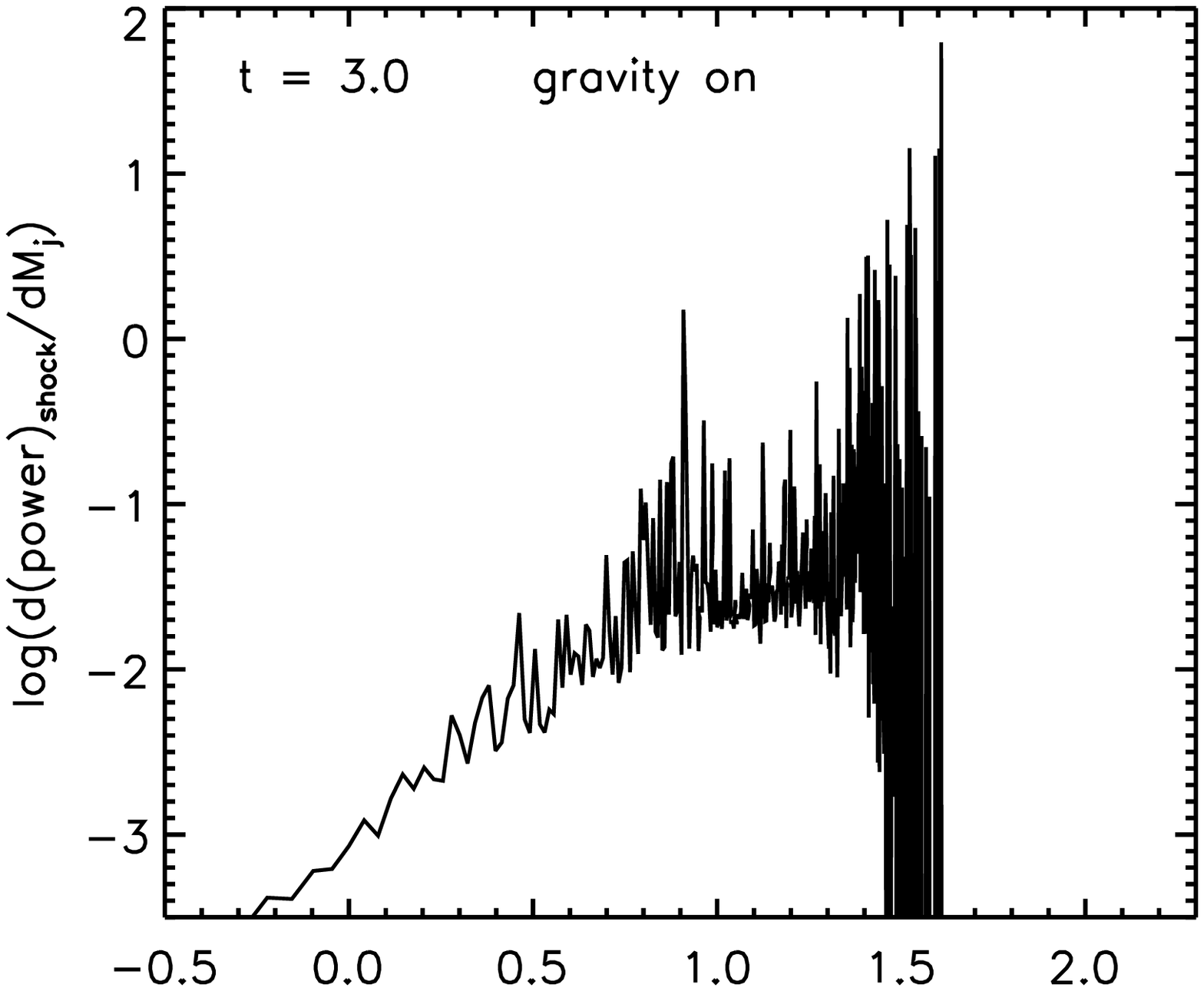,width=0.50\textwidth,angle=0}
    \psfig{file=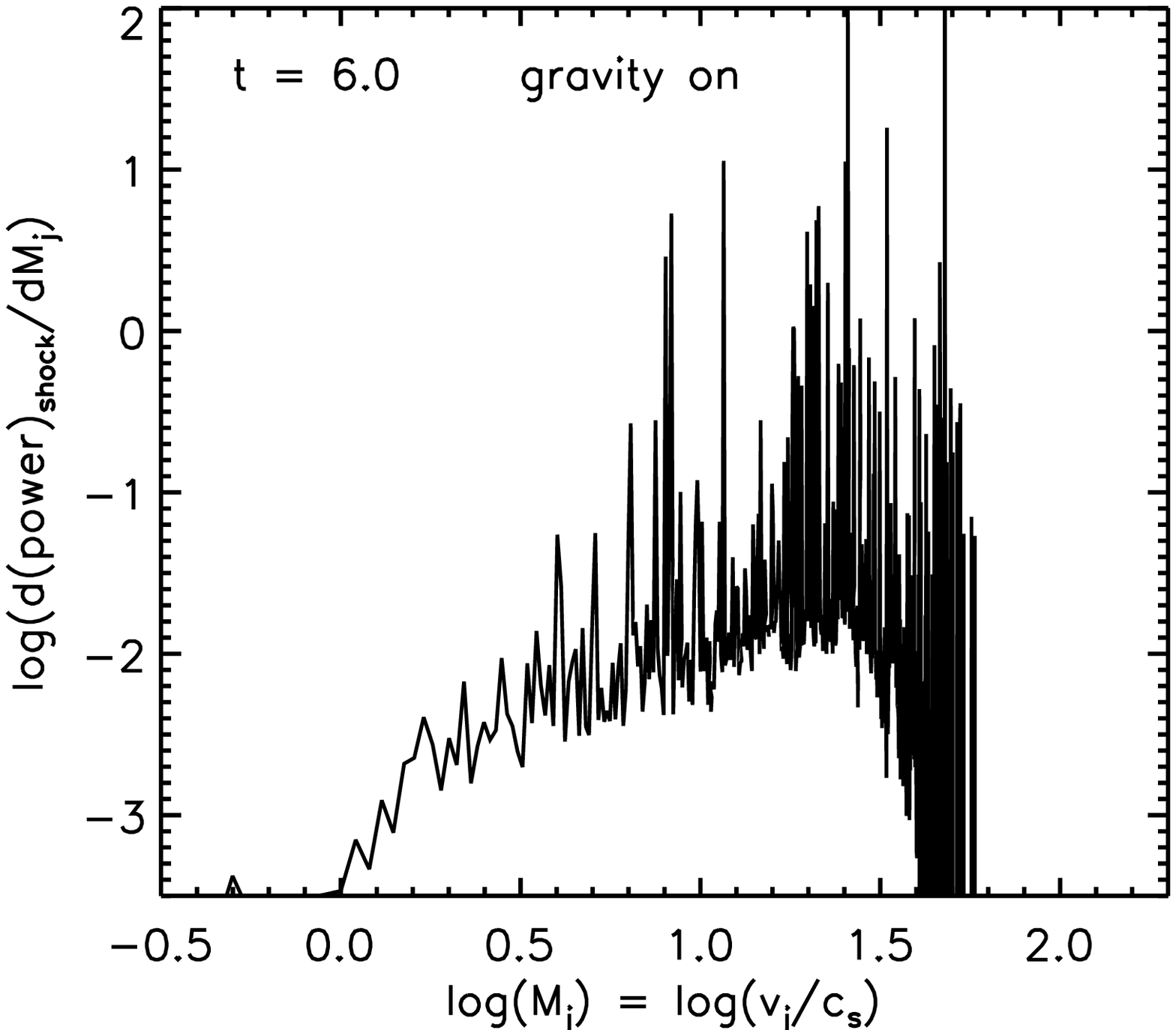,width=0.50\textwidth,angle=0}
\caption{The power dissipated by shocks in driven turbulence with gravity 
at the four indicated times. Gravity is just switched on at the time 
$t\,=\,2$. The parameters are as in Fig\,\ref{selfg}.}   
\label{selfgpower}
  \end{center}
\end{figure}

In stark contrast, the dissipated energy transforms from the typical
non-gravity case (upper box of Fig.\,\ref{selfgpower}) to one dominated 
by a few accretion shocks (lower box of Fig.\,\ref{selfgpower}). A 
relatively small number of well-defined shocks generate strong individual
signatures. These shocks are of typical cloud speeds but are 
strongly dissipative because they propagate into very dense regions. 

\begin{figure*}
  \begin{center}
    \leavevmode
    \psfig{file=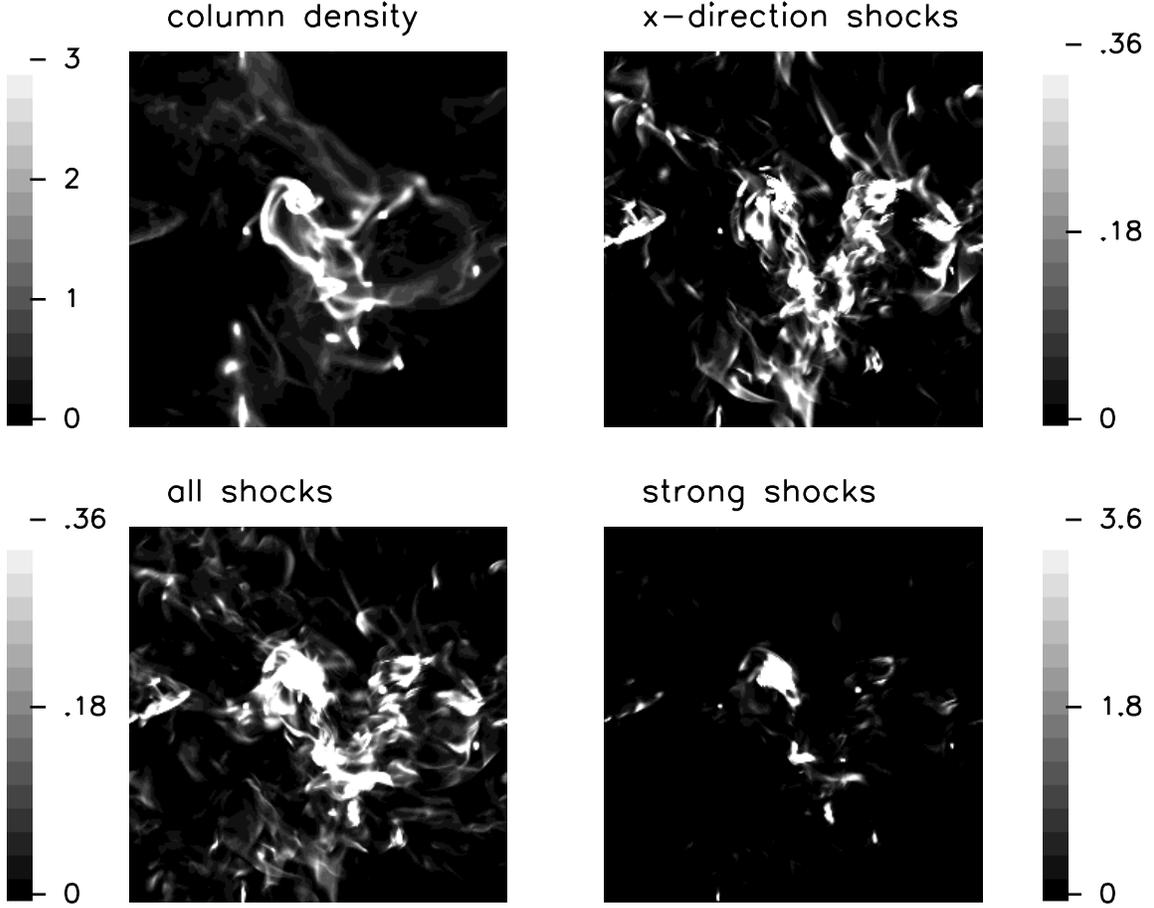,width=0.9\textwidth,angle=0}
\caption{Maps of the column and power dissipated 
in driven turbulence with gravity at the time t\,=\,6. Note the grey
scaling allow a direct comparison with Fig.\,\ref{image-dr}. The
elongated cloud at time  t\,=\,2 has now collapsed into narrow
filaments and cores.}   
\label{image-sg}
  \end{center}
\end{figure*}

The distribution of the shocks at time t\,=\,6 is displayed in 
Fig.\,\ref{image-sg}.
The elongated cloud at time  t\,=\,2  (Fig.\,\ref{image-dr}) has now 
collapsed into narrow filaments and cores. There is high shock dissipation 
around the collapsing cores although other strong shocks are still quite 
widespread. The code does not distinguish between a shock and a collapsing 
flow. The artificial viscosity prescription treats all converging flow 
regions as dissipative. Nevertheless, a collapsing flow does 
physically dissipate its energy as it shocks onto a growing 
core. Hence the dynamical evolution is correctly modelled. The
surface brightness of the shocks is, of course, limited by the grid 
resolution. The high SPH resolution, however, verifies that the 
general properties are realistic (Klessen et al. 2000). A new problem 
arising in the shock analysis  is that a single converging flow onto 
a core may actually be a double shock layer. We would here record 
this as a single shock in the counting procedure and in 
Fig.\,\ref{selfgpower}. We have not tried to correct for this.

\section{Interpretation}

Turbulence has proven easy to interpret, but the interpretations have been
hard to prove. Our aim here is limited to answer: why a universal inverse 
square root power law? Supersonic turbulence possesses simplifying
characteristics which we here employ to understand the shock distribution.

\begin{figure}
  \begin{center}
    \leavevmode
    \psfig{file=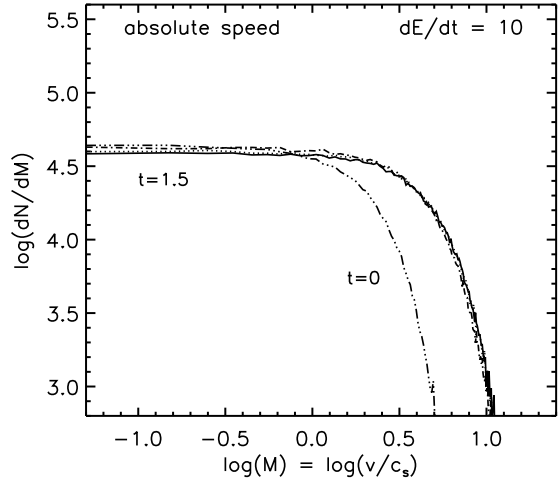,width=0.56\textwidth,angle=0}
\caption{The remarkable number distribution of shock absolute speeds 
in uniform driven turbulence for the four times from t=0 to t=1.5.
The absolute speed is defined as the average speed in the
x-direction within a  region of converging flow along the
x-axis. So defined, we do not have to contend with the shock direction
or with determining the 3-D shock structure, but retain the
vital information for modelling. }   
\label{absolute}
  \end{center}
\end{figure}

First, it is significant that the flow is dominated by the 
integrated number and power of the high-speed shocks at the
knee in the shock jump distributions. These shocks are generated
by the driver and propagate at high speed into the fluid. In contrast,
a shock from the
power-law section plays a passive r\^ole: when overtaken by a
strong shock, its strength decays according to the speed and density
of the oncoming material. Hence the mean lifetime, $\tau$, of
a shock layer, and hence of the shock momentum, is a
constant, independent of its own absolute speed, $v$. Expressed
mathematically, for a steady state, this yields
\begin{equation}
\frac{\partial}{\partial v} \left[\frac{\partial N}{\partial v}
         \frac{1}{\tau}\right] = 0,
\label{interp}
\end{equation}   
from which we find the distribution of absolute shock speeds
(i.e. the speed of the layers within the box) is 
${\partial N}/{\partial v} = constant$. This is confirmed 
from the simulations - Fig.\,\ref{absolute} shows that, below
the high-speed knee, the distribution of absolute velocities 
within the shocks is indeed independent of the absolute speed.

The relationship between the absolute speed, $v$, and the jump speed, $v_j$,
for the ensemble of shocks is expected to resemble that of
Burger's turbulence (Paper 1, see also Gotoh \& Kraichnan 1993). The
steepening of these independently-moving shocks is proportional
to the absolute speed simply because higher amplitude waves produce
stronger shocks, whereas low amplitude waves can only lead to
weak jumps. Secondly, the higher amplitudes imply that the
velocity gradients are also steeper. Hence the material swept up
into a shock is from a wider region which possesses a larger
velocity difference. These two linear effects lead to a relationship
for the ensemble shock jump and absolute speeds of the form
$v_j \propto v^2$. This is the essence of mapping closure
theory, which then uses this relationship to transform initial 
Gaussian distributions into exponentials (Chen et al 1989, Kraichnan 
1990). Here, we transform the  distribution of absolute shock speeds 
generated by the uniform driving, to obtain  
\begin{equation}
    \frac{\partial N}{\partial v_j} \propto v_j^{-1/2}.
\label{interp}
\end{equation} 
This agrees with the numerical experiments. Note that the result depends
on the nature of the driving: uniform driving produces strong shocks
locally which then propagate through the ambient medium. In contrast,
a white-noise driver would  continually create weak small-scale 
turbulence throughout the space and might thus produce, for example, some 
other power-law behaviour.

We found here that the high-velocity tail is somewhat steeper than Gaussian.
A similar result has been obtained for other types of turbulence, with 
the same form occurring, $\exp (-v_j^3)$, as uncovered here (see Gotoh \& 
Kraichnen 1998). This is due to the Gaussian forcing being modified by the high
dissipation rates at these speeds.

The total energy `radiated' in the shocks can be found on integrating
Eq.\,(\ref{eqnpower}) over the jump Mach Number. This yields
$P = 0.669\,\dot E$. The remaining injected energy is also lost 
in the shocks where numerical diffusion is, of course, strong.
As shown in Table 1 of Paper 1, we expect approximately 0.68
of the energy to be radiated in the shocks. Hence, the description
of driven turbulence is fully consistent.

\section{Conclusions}

We have analysed numerical simulations of uniformly driven supersonic, 
magnetohydrodynamic and self-gravitating turbulence. Below a critical jump 
speed, we find  a power law distribution of fast shocks with the number of 
shocks inversely proportional to the square root of the shock jump speed. 
Hence, unlike the decaying case, the driven case possesses an 
`inertial range' of shock strengths. This range is, however, dynamically 
passive, being  mediated by strong shocks injected at a higher knee in the 
distribution.  The knee is mainly responsible for the dissipation of 
energy, and thus the power-law range may only have a weak observational 
signature. This will be explored in the following paper of this series.

These results contrast with the exponential distribution and slow shock 
dissipation associated with decaying turbulence described in Paper 1.  

A strong magnetic field does not alter the shape of the shock number 
distributions. It enhances the shock number transverse to the field
direction at the expense of parallel shocks. The distribution of
parallel shocks is, however, extended to higher jump speeds.
Transverse waves thus appear to find support in the magnetic pressure.

A simulation with self-gravity demonstrates the development of a  number 
of highly dissipative accretion shocks. These  shocks are rare 
and are not apparent in the number distributions. The shock number 
distribution demonstrates a  gradual curvature rather than a power-law.
This implies that high-speed shocks can be maintained by gravitational
acceleration.

Finally, we have shown how the power-law behaviour may arise. No substantial 
theory has been developed to predict this inverse square root law. The 
power law is closely related to the distribution in absolute shock speeds, 
which is shown to be flat. The particular power law is predicted to 
depend on the type of driver. 

The energy is dissipated in shocks mainly associated with the denser
regions of the clouds formed by the large scale wave motions. Numerous
other shocked regions are scattered throughout. Very strong shocks are 
associated with the collapsing cores in the self-gravitating case.

Our results must be considered within the context of the imposed
models. While turbulence has been pinned down in many studies,
further studies have simply extended the number of pins required.
Extensions to this work include: (1) the shock distributions 
corresponding to non-uniform drivers, (2) the spatial propagation 
of turbulent energy and (3) the observable signatures: the spectral 
energy distributions derived from the shock distributions,
\acknowledgements

MDS benefitted greatly from the hospitality of the Max-Planck-Institut f\"ur
Astronomie.  We thank R. Klessen and E. Zweibel for advice and discussions. 
Computations were performed at the MPG Rechenzentrum Garching. M-MML holds a 
CAREER fellowship from the US National Science Foundation, grant number 
AST 99-85392.

\end{document}